\documentclass[10pt]{amsart}
\UseRawInputEncoding
\usepackage[utf8]{inputenc}
\usepackage{amsfonts}
\usepackage{amssymb}
\usepackage{amsmath,amsthm}
\usepackage{amscd}
\usepackage{graphics}
\usepackage{cancel}
\usepackage{pdfsync}
\usepackage[british]{babel} 
\usepackage{mathrsfs}
\usepackage{enumitem}
\usepackage{stmaryrd}
\usepackage{mathrsfs}
\usepackage{wasysym}
\usepackage{latexsym}
\usepackage[colorlinks,linkcolor=blue,citecolor=blue,urlcolor=blue]{hyperref}

\numberwithin{equation}{section}

\newcommand{\beq}{\begin{equation}}
\newcommand{\eeq}{\end{equation}}
\newcommand{\bea}{\begin{eqnarray}}
\newcommand{\eea}{\end{eqnarray}}
\newcommand{\nn}{\nonumber}

\newcommand{\tbf}{\textbf}

\newcommand{\bk}{\begin{cases}}
\newcommand{\ek}{\end{cases}}

\newcommand\ri{\mathrm{i}}

\newtheorem{definition}{Definition}

\theoremstyle{definition}

\newtheorem{remark}{\textbf{Remark}}

\usepackage{enumerate}
\usepackage{float}
\usepackage{cancel}
\usepackage{hyperref}
\usepackage{mathtools}
\usepackage{afterpage}

\allowdisplaybreaks

\begin{document}

\title[A new family of n-dimensional Hamiltonian models]{On higher-dimensional superintegrable systems: \\ A new family of  classical and quantum Hamiltonian models}

\author{Miguel A. Rodr\'iguez}
\address{Departamento de F\'{\i}sica Te\'{o}rica, Facultad de Ciencias F\'{\i}sicas, Universidad
Complutense de Madrid, Plaza de las Ciencias, 1, 28040 -- Madrid, Spain}
\email{rodrigue@fis.ucm.es}
\author{Piergiulio Tempesta}
\address{Departamento de F\'{\i}sica Te\'{o}rica, Facultad de Ciencias F\'{\i}sicas, Universidad
Complutense de Madrid, Plaza de las Ciencias, 1, 28040 -- Madrid, Spain \\  and Instituto de Ciencias Matem\'aticas, C/ Nicol\'as Cabrera, No 13--15, 28049 Madrid, Spain}
\email{piergiulio.tempesta@icmat.es, ptempest@ucm.es}

\date{December 20, 2022}


\begin{abstract}
We introduce a family of $n$-dimensional Hamiltonian systems which, contain, as special reductions, several superintegrable systems as the Tremblay-Turbiner-Winternitz system, a generalized Kepler potential and the anisotropic harmonic oscillator with Rosochatius terms. We conjecture that there exist special values in the space of parameters, apart from those leading to known cases, for which this new Hamiltonian family is superintegrable.

\end{abstract}

\maketitle

\tableofcontents

\section{Introduction}
\subsection{Classical and quantum superintegrability}
In the last two decades, the theory of superintegrable systems has been one of the most active research areas in mathematical physics \cite{TWH04,MPW2013}. These systems represent a special class of integrable systems which possess, both in the classical and in the quantum case, surprisingly rich algebraic and geometric properties  \cite{FMSUW}, \cite{MSVW}, \cite{WSUF}, \cite{Nekh}, \cite{MF},  \cite{STW01}, \cite{TTW01}. Essentially, they  possess more independent integrals of motion than degrees of freedom.  If a system with  $n$ degrees of freedom admits $n+1$  functionally independent integrals of motion, then it is said to be  \textit{minimally superintegrable}; if it admits $2n-1$ integrals (the maximal number allowed), it is \textit{maximally superintegrable}. 

In classical mechanics, as a consequence of Nekhoroshev's theorem \cite{Nekh}, almost all bounded orbits of a maximally superintegrable system are closed and periodic. 

In quantum mechanics, superintegrability is related with the existence of hidden symmetries, responsible for the accidental degeneracy of the spectrum. 

Besides, the notion of maximal superintegrability is intimately related with that of \textit{exact solvability}. An exactly solvable quantum system is a model whose spectrum and eigenfunctions can be obtained by purely algebraic means. It is characterized by the fact that there exists 
in its solution space an infinite flag of functional linear spaces which are preserved by the Hamiltonian.

A particular, well-known example of these flags is given by
finite-dimensional representation spaces of semi-simple Lie algebras of first order differential
operators. Thus, the Hamiltonian and its quantum integrals of motion turn out to be an element of the universal enveloping algebra of a Lie algebra.

Instead, for a quasi-exactly-solvable system, only few energy levels and eigenfunctions can be obtained algebraically.

A conjecture proposed in \cite{TTW01} states that maximally superintegrable quantum systems are exactly solvable.

Another important issue is that of \textit{separability}, namely the construction of suitable coordinate systems (the \textit{separation variables}) in which dynamics decouple. The formulation of algebraic and geometric techniques allowing one to separate variables, both for classical and quantum models, has been historically one of the most relevant problems in the theory of integrable systems, and it has been recently extended to the case of superintegrable models \cite{KKM2018Book}. Typically, superintegrable systems are \textit{multiseparable}, namely separable in more than one coordinate system \cite{RTT2022CNS}. However, this property is not necessary: interestingly enough, superintegrable systems, even maximally ones, may or may not be separable \cite{MPW2013}. A paradigmatic example is the classical Post-Winternitz system \cite{PW2011}, which is indeed maximally superintegrable. However, its Hamilton-Jacobi equation is not additively separable, at least in any orthogonal coordinate system in the configuration space, since the system does not possess integrals of motion quadratic in the momenta, but only higher-degree ones.
 The same is true for the associated Schr\"odinger equation, which does not admit multiplicative separability in any coordinate system (See Theorem 1 of \cite{PW2011}).
On the other hand, there are systems possessing higher-order integrals, which separate in specific coordinate systems \cite{EWYJPA2018}, \cite{PW2015}.

The purpose of this Letter 
is to introduce in a simple way a new (to the best of our knowledge) $n$-dimensional generalization of a class of well-known maximally superintegrable systems. 

\subsection{The new model}

Precisely, we propose the following

\begin{definition} We introduce the Hamiltonian
\begin{equation} \label{RT}
\begin{small}
 H^{(cl)}_n:=(x_1^2+\cdots + x_{n}^2+\gamma)^{\frac{k-1}{k}}\left\{\frac12\sum_{i=1}^{n}p_{x_i}^2 +\frac{1}{2k^2}(\omega_{1}^{2} x_{1}^2+\cdots+ \omega_{n}^{2} x_{n}^2)^{\frac{s-k+1}{k}}+  \frac{\alpha_1  }{ x_{1}^2}+\cdots+\frac{\alpha_n  }{ x_{n}^{2}}\right\} \ ,
 \end{small}
\end{equation}
where  $\gamma\geq0$, $k\neq0$, $s$,  $\alpha_1,\ldots$,$\alpha_n$, $\omega_1,\ldots,\omega_n$ are parameters.
\end{definition}

Our motivation is the fact that, surprisingly, many well-known superintegrable models can be obtained as particular cases of the new family \eqref{RT}. 

We start our analysis by considering first the Hamiltonian \eqref{RT} for $\gamma=0$. 

As we shall prove, for $n=2$, $s=1$ and $\omega_1=\omega_2$ our Hamiltonian \eqref{RT} is equivalent to the well-known Tremblay-Turbiner-Winternitz (TTW) system \cite{TTW09,TTW10}, which has attracted considerable attention in the last decade.  The authors conjectured that the system is maximally superintegrable for all rational values of the parameter $k$. The conjecture was proved for odd $k$ in \cite{Quesne10} and subsequently for all rational values of $k$ in the papers \cite{KKM2010JPA,KMP2011PAN}. 

In order to show the direct relation of system \eqref{RT} with the TTW system, in the next section we shall construct a transformation which leads the TTW system to a new, Cartesian form.

For $n=2$, $s=-1/2$ and  $\omega_1=\omega_2$, the Hamiltonian \eqref{RT} reproduces the generalized Kepler system \cite{PW2010} which is related with the TTW system via the coupling constant metamorphosis mechanism and is  maximally superintegrable.

Another notable reduction of our Hamiltonian, again for $n=2$, is obtained for $k=1$, $s=1$, $\omega_1=\omega_2$: indeed, in this case we get the system Smorodinsky-Winternitz I \cite{WSUF}, which is one of the four superintegrable systems in the plane with quadratic integrals of motion. Also, for $k=1$, $s=1$, and $\omega_1=2\omega$, $\omega_2= \omega$ we obtain the Smorodinsky-Winternitz system II.

\subsection{The n-dimensional reductions}
The Hamiltonian \eqref{RT} also admits several interesting reductions to $n$-dimensional superintegrable models.
First, we observe that we can reproduce both the classical harmonic potential and the Kepler potentials in all dimensions. 
Besides, for $k=1$, $s=1$, $\omega_1=\ldots=\omega_n$, one readily obtains from system \eqref{RT} the $n$-dimensional isotropic harmonic oscillator with Rosochatius terms \cite{Rosochatius}. It is also a maximally superintegrable system, as has been proved independently in \cite{EV08} and \cite{RTW08}.

A crucial aspect is that system \eqref{RT} contains, as particular reductions, $n$-dimensional versions of both the TTW system and the generalized Kepler system. These two systems are defined in the Euclidean plane in terms of trigonometric functions, and (to the best of our knowledge) no generalizations to $n$ dimensions of them were known. Precisely, for sake of concreteness, our generalization of the TTW system reads:
\begin{equation} \label{nTTW}
\begin{small}
 H^{(gTTW)}_n:=(x_1^2+\cdots + x_{n}^2)^{\frac{k-1}{k}}\left\{\frac12\sum_{i=1}^{n}p_{x_i}^2 +\frac{\omega^{2}}{2k^2}( x_{1}^2+\cdots+ x_{n}^2)^{\frac{2-k}{k}}+  \frac{\alpha_1  }{ x_{1}^2}+\cdots+\frac{\alpha_n  }{ x_{n}^{2}}\right\} . 
 \end{small}
\end{equation}
We propose the following system, that we call the  $n$-dimensional generalized Kepler system:

\begin{equation} \label{nPW}
\begin{small}
 H^{(gKs)}_n:=(x_1^2+\cdots + x_{n}^2)^{\frac{k-1}{k}}\left\{\frac12\sum_{i=1}^{n}p_{x_i}^2 -\frac{\omega^2}{2k^2}( x_{1}^2+\cdots+ x_{n}^2)^{\frac{1-2k}{2k}}+  \frac{\alpha_1  }{ x_{1}^2}+\cdots+\frac{\alpha_n  }{ x_{n}^{2}}\right\} \ .
 \end{small}
\end{equation}

In other words, both the TTW and the PW systems come from Hamiltonian \eqref{RT} for $\gamma=0$, $n=2$, once written in suitable polar coordinates, assuming equal  frequencies (imaginary for the gKs system) and choosing $s=1$ and $s=-1/2$, respectively.
Due to the physical relevance of the three-dimensional case, we have performed a numerical study of the generalized Kepler system \eqref{nPW}. Our analysis strongly suggests that for $n=3$ and $n=4$ the gKs is maximally superintegrable.

The Hamiltonian \eqref{RT} is also related with the interesting family of Hamiltonians studied by Ballesteros, Enciso, Herranz and Ragnisco in \cite{BEHRAOP2009}. This general family describes the motion of a particle on any $n$-dimensional spherically symmetric
curved space and is quasi-maximally superintegrable, namely it possesses $2n-2$ integrals for any choice of the two arbitrary functions of $r$ appearing in it. Precisely, the particular case of Hamiltonian \eqref{RT} with $\omega_1=\ldots=\omega_n$ and $s=-1$ falls into this class, when the central potential $\mathcal{U}(r)$ vanishes.  

Finally, for $n$ arbitrary, $\gamma>0$, $k=1/2$, $s=0$, $\omega_1=\ldots=\omega_n$ we recover the class of maximally superintegrable models on an $n$-dimensional space of nonconstant curvature, introduced in \cite{BEHRPD2008}.

\vspace{3mm}

A quantum version of Hamiltonian \eqref{RT}, namely the system \eqref{qRT}, will also be presented and its properties briefly discussed.  We  conjecture that for special values of the space of parameters of the quantum system \eqref{qRT}, we will obtain exactly solvable models. 

Clearly, the issue of integrability  of our family \eqref{RT} for arbitrary values of the parameters, both for the classical and quantum cases, is a very interesting one. Some related conjectures and open problems will be stated in the final section. 

The problem of separability of system \eqref{RT} is also completely open, although there are certainly separable subcases. For instance, the TTW system is separable in polar coordinates for any value of $k$. 

\vspace{3mm}

The paper is organized as follows. In Section 2, we formulate in new variables the standard TTW system. In this way, the link with the family of systems \eqref{RT} becomes transparent. Also, we propose our generalization of the original TTW system by means of an anisotropic deformation of the harmonic term and by an extra parameter interpolating between the Kepler and the anisotropic cases.
In Section 3, we introduce a quantum version of our Hamiltonian. We propose several open problems in the final Section 4.

\section{On the construction of the new Hamiltonian family}

The classical version of the TTW Hamiltonian in polar coordinates reads:
\begin{equation} \label{pTTW}
H^{TTW}=\frac{1}{2}\bigg(p^{2}_{r}+\frac{p^{2}_{\varphi}}{r^2}\bigg) +\frac{1}{2}\omega^2 r^2+\frac{\alpha k^2}{2r^2\cos^2k\varphi}+\frac{\beta k^2}{2r^2\sin^2k\varphi}.
\end{equation}
We shall prove that for $n=2$, $s=1$, $\gamma=0$ and $\omega_1=\ldots=\omega_n=\omega$ Hamiltonian \eqref{RT} reduces to system \eqref{pTTW}. 
To this aim, first we shall rewrite Hamiltonian \eqref{pTTW} in Cartesian coordinates. We get
\begin{equation} \label{2.2}
{H}_{cart}^{TTW}:=\frac12(p_x^2+p_y^2)+\frac12\omega^2(x^2+y^2)+V_k(x,y;\alpha,\beta),
\end{equation}
where we have introduced the potential 
\begin{equation}
V_k(x,y;\alpha,\beta)=2k^2(x^2+y^2)^{k-1}\bigg(\frac{\alpha }{((x+\ri y)^k+(x-\ri y)^k)^2}-\frac{\beta }{((x+\ri y)^k-(x-\ri y)^k)^2}\bigg)
.
\end{equation}

Let us introduce two complex variables:
\begin{equation} \label{comp1}
 z=x+\ri y,\quad p_z=\frac12(p_x-\ri p_y)
\end{equation}
and their complex conjugates 
\begin{equation} \label{comp2}
\bar{z}=x-\ri y,\quad p_{\bar{z}}=\frac12(p_x+\ri p_y).
\end{equation}
This change of variables preserves the canonical symplectic form in $\mathbb{R}^4$.
The TTW Hamiltonian can be written in these new coordinates as:
\begin{equation}\label{compham}
 \mathfrak{H}^{TTW}:=2 p_zp_{\bar{z}}+\frac12\omega^2z\bar{z}+V_k(z,\bar{z};\alpha,\beta).
\end{equation}
Using the de Moivre formula:
\begin{equation}
z=r^{\ri\varphi},\quad z^k=r^k(\cos k\varphi+\ri\sin k\varphi)
\end{equation}
that is
\begin{equation}
\cos k\varphi=\frac{1}{2|z|^{k}}(z^k+\bar{z}^k),\quad \sin k\varphi=\frac{1}{2\ri|z|^{k}}(z^k-\bar{z}^k)
\end{equation}
we can express the potential as
\begin{equation}\label{compot}
V_k(z,\bar{z};\alpha,\beta)=\frac{2\alpha k^2z^{k-1}\bar{z}^{k-1}}{(z^k+\bar{z}^k)^2}-\frac{2\beta k^2z^{k-1}\bar{z}^{k-1}}{(z^k-\bar{z}^k)^2} \ .
\end{equation}

If $k=1$ we recover the harmonic oscillator with centrifugal terms:
\begin{equation}
V_1(z,\bar{z};\alpha,\beta)=\frac{2\alpha }{(z+\bar{z})^2}-\frac{2\beta }{(z-\bar{z})^2}
\end{equation}
and, in fact, \eqref{compot} is the general form of the potential for any $k$.
We can make another change of coordinates, in this case to real coordinates:
\begin{equation}\label{sec1}
u=\frac12(z^k+\bar{z}^k),\quad v=\frac{1}{2\ri}(z^k-\bar{z}^k)
\end{equation}
yielding the corresponding transformation for the momenta
\begin{equation} \label{sec2}
p_z=\frac{k}{2}z^{k-1}(p_u-\ri p_v),\quad p_{\bar{z}}=\frac{k}{2}\bar{z}^{k-1}(p_u+\ri p_v) \ .
\end{equation}
Consequently, the Hamiltonian \eqref{compham} in these new \textit{Cartesian} coordinates reads
\begin{equation}\label{conflat}
\mathfrak{H}_k=k^2(u^2+v^2)^{\frac{k-1}{k}}\bigg[\frac12(p_u^2+ p_v^2)+\frac{\omega^2}{2k^2}(u^2+v^2)^{\frac{2}{k}-1}+\frac{\alpha}{2u^2}+\frac{\beta}{2v^2}\bigg] \ .
\end{equation}
\begin{remark}
The global factor in eq. \eqref{conflat} can be interpreted as a conformally flat metric; the remaining Hamiltonian factor represents a nonlinear oscillator with centrifugal barriers. These barriers can also arise from the reduction of a nonlinear oscillator in a four dimensional space \cite{RTW08,EV08}; alternatively, they can be obtained from the construction given in \cite{RR10}. 
\end{remark}

Let us now introduce a new parameter $s$, which allows us to treat the Kepler and the harmonic cases on the same footing. Precisely, \begin{equation} 
\mathfrak{H}_k^{(s)}(x,y)=\frac12(p_x^2+p_y^2)+\frac12\omega^2(x^2+y^2)^s+V_k(x,y;\alpha,\beta)
\end{equation}
 Using \eqref{comp1} and \eqref{comp2}, we get
\begin{equation} 
 \mathfrak{H}^{(s)}_k=2 p_zp_{\bar{z}}+\frac12\omega^2(z\bar{z})^s+V_k(z,\bar{z};\alpha,\beta)
\end{equation}
As before, taking into account eqs. \eqref{sec1} and \eqref{sec2} we get
\begin{equation} 
(z\bar{z})^{s}= (u^2+ v^2)^{s/k},\quad p_zp_{\bar{z}}=\frac{k^2}{4}(z\bar{z})^{k-1}(p_u^2+p_v^2)
\end{equation}
and
\begin{equation} 
 V_k(u,v;\alpha,\beta)=\frac{k^2}{2}(u^2+v^2)^{(k-1)/k}\left(\frac{\alpha  }{ u^2}+\frac{\beta  }{ v^2}\right).
\end{equation}
Thus, we obtain the new Hamiltonian
\begin{equation} 
 \mathfrak{H}_k^{(s)}=\frac{k^2}{2}(u^2+ v^2)^{ (k-1)/k}(p_u^2+p_v^2) +\frac12\omega^2(u^2+ v^2)^{s/k}+\frac{k^2}{2}(u^2+v^2)^{ (k-1)/k}\left(\frac{\alpha  }{ u^2}+\frac{\beta  }{ v^2}\right)
\end{equation}
or
\begin{equation} \label{sconflat}
 \mathfrak{H}_k^{(s)}= k^2 (u^2+ v^2)^{(k-1)/k}\left\{\frac12(p_u^2+p_v^2) +\frac{ \omega^2}{2k^2}(u^2+ v^2)^{(s-k+1)/k}+ \frac{ \alpha  }{ 2u^2}+\frac{ \beta  }{ 2v^2}\right\}.
\end{equation}

\begin{remark}
The form of the Hamiltonian \eqref{sconflat} was the inspiration source for our Hamiltonian \eqref{RT}. This system is more general than system \eqref{sconflat} for two reasons. The first one is that our Hamiltonian \eqref{RT} is defined in \textit{$n$ dimensions}. The second one is that we allow for anisotropy in the oscillations along distinct axes. This, in turn, relates directly the Hamiltonian \eqref{RT} with well-known anisotropic oscillators. 

Finally, we observe that the new parameter $s$ allows us to continuously interpolate between the harmonic and Kepler cases.
\end{remark}

\section{The quantum version of the model}

We wish to propose now the quantum analogue of our Hamiltonian models.
\begin{definition}
We introduce the family of quantum Hamiltonians
\begin{eqnarray} \label{qRT}
\nn  H^{(q)}_n&:=& (x_1^2+\cdots+ x_{n}^2+\gamma)^{\frac{k-1}{k}} \cdot \\
\nn && \left\{-\frac12(\partial_{x_1}^2+\cdots+ \partial_{x_n}^2) +\frac{1}{2 k^2}(\omega_{1}^{2} x_{1}^2+\cdots + \omega_{n}^{2} x_{n}^2)^{\frac{s-k+1}{k}}+  \frac{\alpha_1}{x_{1}^2}+\cdots+\frac{\alpha_n}{ x_n^{2}}\right\} \ ,
\\
\end{eqnarray}
where $\gamma\geq 0$, $k\neq0$, $s$, $\alpha_i$,  $\omega_i$, $i=1,\ldots,n$ are parameters.
\end{definition}
This quantum Hamiltonian has several interesting reductions. For sake of simplicity, we shall consider $\gamma=0$.

\begin{itemize}
\item For $k=1$, $s=1$ it reproduces a quantum anisotropic oscillator with Rosochatius terms, which is maximally superintegrable when all frequencies coincide. 

\item For $k=1$, $s=-1/2$ and equal frequencies we get again a Kepler potential with a Rosochatius deformation.  

\item For $k=1$, $s=-1/2$, equal frequencies and $\alpha_n=0$ we obtain the system 
\beq
H =  -\frac12(\partial_{x_1}^2+\cdots+ \partial_{x_n}^2) -\frac{\lambda}{r}+  \frac{\alpha_1}{x_{1}^2}+\cdots+\frac{\alpha_{n-1}}{ x_{n-1}^{2}} \ ,
\eeq 
with $r=\sqrt{x_1^2+\cdots x_n^2}$, which was proved in \cite{RW02} to be exactly solvable.

\item The two dimensional case
\begin{equation} \label{qRT}
 H^{(q)}_2:=(x^2+ y^2)^{\frac{k-1}{k}}\left\{-\frac12(\partial_{x}^2+\partial_{y}^2) +\frac{1}{2 k^2}(\omega_{1}^{2} x^2+\omega_2^2 y^2)^{\frac{s-k+1}{k}}+  \frac{\alpha}{x^2}+\frac{\beta}{y^2}\right\} \ ,
\end{equation}
is quite interesting by itself. Notice that for $k=1$, $s=1$, and $\omega_1=2\omega$, $\omega_2= \omega$, $\alpha=0$ we recover the maximally superintegrable and exactly solvable  Smorodinski-Winternitz system II \cite{MSVW,STW01,TTW01}:
\begin{equation}
H^{SWII}=-\frac12(\partial_{x}^2+\partial_{y}^2) + \frac{\omega^{2}}{2}(4 x^2+y^2)+  \frac{\beta}{y^2}
\end{equation}
once we annul the constant multiplying the linear term in $x$.

\vspace{3mm}
 
\item Again in the case $n=2$, and for $\omega_1=\omega_2$, $s=1$, Hamiltonian \eqref{qRT} reduces to the quantum TTW system \cite{TTW09}, whereas, for $s=-1/2$, it reduces to the quantum version of the generalized Kepler model \cite{PW2010}. This can be verified by means of variable transformations completely analogous to those illustrated in Section 2.

\item In three dimensions, a quantum version of the TTW system, in spherical coordinates, has been proposed in \cite{KKM2010JPA}. It differs from the one that we obtain by specializing formula \eqref{nPW} for $n=3$.
\end{itemize}

\section{Open problems and conjectures}

The purpose of this Letter was, in essence, to introduce the Hamiltonian \eqref{RT}. It represents a large family of Hamiltonian systems, including as particular cases or reductions several well-known superintegrable models that attracted much attention in the last decades. Many questions concerning Hamiltonian \eqref{RT} remain open.
Thus, we propose the following, general problems.

\vspace{3mm}

\textbf{Problem 1}. Find all the values of the parameters appearing in the classical Hamiltonian \eqref{RT} which make it integrable or (maximally) superintegrable, in all dimensions. Determine the orbits of the motion.

\vspace{3mm}

\textbf{Problem 2}. Find the values of the parameters appearing in the quantum Hamiltonian \eqref{qRT} which make it integrable or superintegrable. Find the spectrum and the associated eigenfunctions. 

\vspace{3mm}

\textbf{Problem 3}. Find, for special values of the parameters, separation variables for the Hamilton-Jacobi equation associated with the classical Hamiltonian \eqref{RT} and for the Schr\"odinger equation of the quantum case \eqref{qRT}.

\vspace{3mm}

\textbf{Problem 4}. Determine whether the quantum Hamiltonian \eqref{qRT} is exactly solvable or quasi-exactly solvable for special values of the parameters.

\vspace{3mm}

We conjecture that there exist new values of the parameters ensuring classical (maximal) superintegrability and quantum exact solvability, apart from those leading to known reductions.

\section*{Acknowledgements}

We wish to thank G. Gubbiotti and D. Latini for interesting discussions. This work has been partly supported by  the Universidad Complutense de Madrid under grant G/6400100/3000, and  by the Severo Ochoa Programme for Centres of Excellence in R\&D
(CEX2019-000904-S), Ministerio de Ciencia, Innovaci\'{o}n y Universidades y Agencia Estatal de Investigaci\'on, Spain.

P. T. is member of the Gruppo Nazionale di Fisica Matematica (GNFM).

\end{document}